\makeatletter\def\graphicscache@inhibit{true}\makeatother
\documentclass[runningheads]{llncs}
\usepackage[T1]{fontenc}
\usepackage{hyperref}
\usepackage{graphicx}
\usepackage{svg}
\usepackage{wrapfig}
\usepackage[inline]{enumitem}
\usepackage{footmisc}
\usepackage[backend=biber,style=lncs,url=true]{biblatex}
\begin{document}
\title{XFSC: A Catalogue of Trustable Semantic Metadata for Data Services and Providers}
\titlerunning{XFSC Catalogue: Trustable Semantic Service and Provider Metadata}
\author{Benedikt T. Arnold\inst{1,2}\orcidID{0000-0001-8594-880X} \and
Khalil Baydoun\inst{1,2} \and
Diego Collarana\inst{1}\orcidID{0000-0002-2583-0778} \and
Sebastian Duda\inst{1}\orcidID{0000-0003-3246-4021} \and
Christina Gillmann\inst{1,2} \orcidID{0009-0005-9087-205X} \and
Ahmad Hemid\inst{1}\orcidID{0000-0002-9811-0579}
Philipp Hertweck\inst{3}\orcidID{0000-0002-1051-8052} \and
Paul Moosmann\inst{1}\orcidID{0009-0005-2114-8578} \and
Denis Sukhoroslov\inst{4} \and
Christoph Lange\inst{1,2}\orcidID{0000-0001-9879-3827}
}
\authorrunning{Arnold et al.}
\institute{Fraunhofer Institute for Applied Information Technology FIT, Germany\\
\email{benedikt.arnold@fit.fraunhofer.de}\\ 
\and
RWTH Aachen University, Information Systems, Germany
\and
Fraunhofer IOSB, Karlsruhe, Germany\\
\and
T-Digital by Deutsche Telekom, Thessaloniki, Greece
}
\maketitle              %
\begin{abstract}
In dataspaces, federation services facilitate key functions such as enabling participating organizations to establish mutual trust and assisting them in discovering data and services available for consumption.
Discovery is enabled by a catalogue, where participants publish metadata describing themselves and their data and service offerings as Verifiable Presentations (VPs), such that other participants may query them.
This paper presents the Eclipse Cross Federation Services Components (XFSC) Catalogue, which originated as a catalogue reference implementation for the Gaia-X federated cloud service architecture but is also generally applicable to metadata required to be trustable.
This implementation provides basic lifecycle management for DCAT-style metadata records and schemas.
It validates submitted VPs for their cryptographic integrity and trustability, and for their conformance to an extensible collection of semantic schemas.
The claims in the latest versions of valid VP submissions are extracted into a searchable graph database. %
The implementation scales to large numbers of records and is secure by design.

Filling the catalogue with content in a maintainable way requires bindings towards where data and service offerings are coming from: connectors that expose resources hosted in an organization's IT infrastructure towards the dataspace.  We demonstrate the integration of our catalogue with the widely used Eclipse Dataspace Components Connector, enabling real-world use cases of the German Culture Dataspace.
In addition, we discuss potential extensions and upcoming integrations of the catalogue.%
\keywords{%
Semantic Data Catalogue \and
Dataspaces \and
Federation Services \and
Gaia-X \and
Eclipse Dataspace Components
}
\end{abstract}

\section{Introduction}
\label{sec:intro}

Modern IT landscapes are characterized by a rising usage of external computing resources.
The importance of available data has been increasing lately, especially with the broadening adoption of AI systems, and data sharing has opened new business models, making the value of data visible. Although the need for sharing data is recognized, some hurdles exist. Fear of losing control of (sensitive) data prevents its exchange. This is addressed by emerging data infrastructure initiatives, in particular dataspaces~\cite{Otto2022}, which focus on sovereign data exchange and data usage control.

In recent years, the adoption of dataspaces is sharply increasing. The ``Common European Data Spaces'' are part of the European Union's data strategy~\cite{common-eu-ds}.
As the EU-funded Data Spaces Support Centre (DSSC) describes it, they promote a ``data sovereign, interoperable and trustworthy data sharing environment, to enable data reuse, %
[$\ldots$] fully respecting EU values, and supporting the European economy and society''~\cite{dssc-mission}.

The discoverability of resources inside a decentralized dataspace ecosystem is limited. Complicating, higher-level services can be composed of multiple basic services operated and hosted in different computing environments and provided by different organisations. Therefore, information about the complete stack required to provide one smart service is scattered. These two properties, \begin{enumerate*}\item  decentralized ecosystem and \item service information provided by different organisations\end{enumerate*}, lead to the need for a searchable directory to find services and data.

This paper presents the Cross Federation Services Components (XFSC\footnote{This open-source project has been initiated as ``Gaia-X Federation Services'' (GXFS) and later rebranded during the transfer into the governance of the Eclipse Foundation (cf.\ Section~\ref{sec:adoption}).}) 
Catalogue as a directory of trustable semantic metadata describing services and their providers. It has been developed as a reference implementation for Gaia-X but is also usable for more generic metadata adhering to a schema and required to provide trustability.

To prove the catalogue's usefulness, we develop bindings towards dataspace software -- populating the catalogue in a maintainable fashion. Concretely, we demonstrate the integration with the broadly adopted Eclipse Dataspace Components Connector (EDC Connector, for brevity: the EDC). This integration is already put into action in an existing dataspace, namely the German Culture Dataspace, a lighthouse project in context of the national Digital Strategy.

Section~\ref{sec:background} provides background information on dataspaces, the EDC Connector and also Gaia-X, followed by an overview of related work in Section~\ref{sec:related}. Section~\ref{sec:architecture} presents the basic concepts for managing semantic metadata in a dataspace's catalogue and the trust mechanisms as well as the concept for the integration of our catalogue with the EDC. Section~\ref{sec:implementation} covers the implementation of this functionality. Section~\ref{sec:limitations} discusses the limitations of the implementation and the underlying concepts. Afterwards, Section~\ref{sec:evaluation} presents an evaluation of the component and Section~\ref{sec:adoption} reviews the adoption so far. Section~\ref{sec:conclusion} concludes our work and gives an outlook to future work.

\section{Background}
\label{sec:background}
\noindent\textbf{Dataspaces:}
According to the DSSC Glossary, a dataspace is an ``interoperable framework, based on common governance principles, standards, practices and enabling services, that enables trusted data transactions between participants''~\cite{dssc-glossary}. %
Dataspaces have the goal to manage and exchange data in domain-specific communities, such as mobility, manufacturing, or research. There are several big dataspace initiatives, such as the International Data Spaces (IDS) Gaia-X, or the Common European Data Spaces, with the latter two particularly focusing on complying with the EU rules and values.
To realize dataspaces, a common, machine-readable and interpretable format for exchanging information about services and their providers is needed~\cite{dssc-sd}, which we call \emph{Self-Descriptions} (SD) here, initially given by the provider.

\noindent\textbf{Gaia-X:}
Gaia-X is a federated and secure data infrastructure designed to foster an ecosystem where services are shared in a trustworthy environment, ensuring data sovereignty~\cite{GaiaX:About}. It consists of multiple domain-specific federations that operate under a shared set of compliance rules and policies defined in the Gaia-X Architecture Document~\cite{GaiaX:ArchitectureDocument2404}. To enhance trust and efficiency, Gaia-X adopts a ``compliance as code'' approach, using formalized rules (e.g., semantic schema verification) implemented in automated verification components. Key entities from Gaia-X's architecture are \emph{Participant}s that offer or consume \emph{ServiceOffering}s, aggregated of multiple \emph{Resources} or other \emph{ServiceOfferings}.

\noindent\textbf{Gaia-X Credentials:}
In Gaia-X, SDs are called \emph{Gaia-X Credentials} and make use of established standards. %
The structure of the contained metadata is given by an ontology and SHACL shapes, building on \emph{DCAT} for resources and the \emph{Organization Ontology} for participants. Gaia-X Credentials~\cite{GaiaX:ICAM} are W3C Verifiable Credentials (VCs)~\cite{W3C:VCDM} serialized in JSON-LD. They aggregate claims about entities stated as RDF triples. These VCs, published as W3C Verifiable Presentations (VPs), form the complete SD.
SDs are issued by Gaia-X participants in a decentralized manner, employing the Self-Sovereign Identity (SSI)~\cite{GaiaX:trustworthyecosystems} paradigm. %
Schema verification for compliance with SHACL shapes and integrity and authenticity verification using cryptographic signatures are conceptually foreseen. %
Semantic interoperability is achieved through the above mentioned ontology, aligned with frameworks like \emph{DCAT} for resources and the \emph{Organization Ontology} for participants. %

\noindent\textbf{Eclipse Dataspace Components Connector (EDC):} A dataspace connector acts as a gateway to the dataspace and facilitates the communication with other participants and federation services. The EDC~\cite{edc} is widely used in real dataspaces and designed to be compliant with both IDS- and Gaia-X-based dataspaces, e.g., in the German Mobility~\cite{mds} or Culture Dataspace~\cite{drk}, but also in the \emph{deployEMDS} project building the EU Mobility Dataspace~\cite{deployemds}. %
However, the in-built catalogue facilities are not as semantically rich as our solution and do not offer a well-shaped centralized search. This motivates the integration of the EDC with the XFSC Catalogue. %

The EDC handles so called \emph{assets}. An asset consists of metadata on some resource and a machine-readable description on how to access the resource, e.g., the URL of a REST API. This asset is combined with a usage policy. This together forms an offer for the dataspace.

\section{Related Work}
\label{sec:related}

\noindent\textbf{Metadata Repositories and Data Catalogues} are databases storing metadata to facilitate its management, and by extension search \& discovery, governance, and collaboration.

Amundsen~\cite{amundsen} aims to answer questions about data availability, trustworthiness, ownership, usage, and reusability. Amundsen's core features include metadata ingestion, search, discovery, lineage, and visualization. Its architecture consists of multiple services, including a metadata service, a search service, a frontend service, and a data builder. These services rely on the Neo4j graph database and the NoSQL database Elasticsearch~\cite{Elasticsearch}.

Apache Atlas~\cite{apache-atlas} is an open-source tool for solving search, discovery, and governance problems in the Hadoop ecosystem. It has a wide range of features, such as metadata management, classification, lineage, search, discovery, security, and data masking, which are powered by actively developed and used technologies like the graph database JanusGraph, the search platform Solr, the streaming platform Kafka, and Ranger, a framework to manage data security .

OpenMetadata~\cite{open-metadata} is a data catalog built by the team behind Uber's data infrastructure. 
The technical architecture of OpenMetadata relies upon PostgreSQL's graph capabilities to store relationships. OpenMetadata works towards metadata centralization to enable governance, quality, profiling, lineage, and collaboration.

\noindent\textbf{Ontology Repositories} serve as platforms and registries for managing and maintaining ontologies. They facilitate discovering, accessing, and reusing existing ontologies. Their functionality to support ontologies can vary based on the application domain, which spans from specific to more generalized contexts. For instance, BioPortal~\cite{BioPortal} and OBO Foundry~\cite{obofoundry}
host biomedical ontologies, whereas Linked Open Vocabularies (LOV~\cite{LOV}) encompasses ontologies from multiple domains. A comprehensive overview of existing ontology repositories can be found in the W3C Wiki~\cite{ontoRep}.

\noindent\textbf{Dataspace Catalogues}
are implementations of data catalogues as services inside dataspaces. Since dataspaces have some specific requirements, e.g., regarding data sovereignty, or the use of semantics, there are dedicated catalogue solutions for dataspaces, %
beyond the more general data catalogues presented before. 

As a part of the Eclipse Dataspace Components introduced in Section~\ref{sec:background}, federated catalogue services have been implemented as dataspace extensions. The goal of these services is to enable publishing and finding data. They consist of a federated cache node, which makes its asset catalogue available to other participants and a federated cache crawler, which periodically crawls other nodes~\cite{FederatedCatalogServicesArchitecture}. 

Sovity offers a Catalog-as-a-Service approach, compatible with the EDC Connector. It also enables crawling of data offerings and advanced searching and filtering functionalities. The data offerings are linked to the relevant data owners, indicating where the data can be contracted~\cite{sovityDS}.

International Data Spaces (IDS~\cite{IDS-RAM4}) is a distributed architecture closely related to Gaia-X, but with a narrower scope focusing on sovereign data exchange.
In any IDS dataspace, a Metadata Broker~\cite{IDSAMetadataBrokerSpec}
serves as a metadata registry of local data sources published by the individual participants' data exchange interfaces, the IDS connectors. 
To search over the metadata, the reference implementation stores them in Apache Jena Fuseki~\cite{Fuseki} and uses Elasticsearch~\cite{Elasticsearch} for indexing and full-text search. 

The Pontus-X Gaia-X Web3 Ecosystem features another implementation of a data catalogue, using the blockchain-based Ocean Protocol~\cite{OceanProtocol} to synchronize data between multiple instances. Searching the catalogue~\cite{pontus-x} is realized by Elasticsearch.
The metadata can be encrypted and further access control applied. The credentials are verified against the Gaia-X Compliance Service; supporting additional schemas is not foreseen.

To federate dataspace catalogues, synchronization between different catalogue implementations is necessary. For Gaia-X compliant dataspaces the Credential Events Service~\cite{ces} (CES) provides a solution. It is a software component that stores and shares Gaia-X Credentials using the event data specification CloudEvents~\cite{cloud-events}. To share the credentials the catalogue must implement a way to push and pull credentials to the CES. The XFSC Catalogue has such an implementation.

\section{Catalogue: a Directory of Trustable Semantic Metadata}%
\label{sec:architecture}

In contrast to earlier approaches, one of the key principles of Gaia-X is the continuous consideration of decentralization. Based on commonly agreed policies and rules, independent federations are emerging. Since a variety of services can be provided by different participants, discoverability of offerings within a federation (and, in future, across federations) is needed.
This is supported by the Catalogue as a directory of SDs, which comes along with the following requirements, as defined in Section 7.3 of the architecture document~\cite{GaiaX:ArchitectureDocument2404}:
\begin{enumerate}
    \item \label{req:a} Compliance with Gaia-X standards, i.e., data structure, SD-lifecycle
    \item \label{req:b} Capability for application/domain specific schemas
    \item \label{req:c} Cryptographic and semantic verification of SDs
    \item \label{req:d} Searchability and comparability of services 
    \item \label{req:e} Integration with actively used dataspace connectors
\end{enumerate}

Thus, the XFSC Catalogue consists of multiple components:
\textbf{Self-Description management}
to store and track SDs through their lifecycle (Section~\ref{subsec:arch:lifecycle}),
\textbf{Schema management} 
to handle the current schemas, terms and rules for validating SDs (Section~\ref{subsec:arch:schema-lifecycle}),
\textbf{Verification}
of syntax, semantics and cryptographic signatures of SDs (Section~\ref{subsec:arch:validation}), and a
\textbf{Query component}: an openCypher interface offering search capabilities over the claims of all SDs (Section~\ref{subsec:arch:query}).
A REST API exposes the functionality, which is structured along the functional components. The complete API specification is available in the public GitLab project~\cite{fc-apis}. In addition to the checks done by the verification component, parts of the interface of the Catalogue are protected by authentication and authorization. 
Role Based Access Control ensures that specific endpoints of the API are only available to specific roles among the following:
\begin{enumerate*}
\item the \textbf{Catalogue administrator} is responsible for managing the schemas in the Catalogue (usually the federator operating the Catalogue), 
\item the \textbf{Participant administrator} is responsible for managing the user accounts related to a participant, and 
\item \textbf{Users} act on behalf of at least one participant and are able to upload SDs for their associated participant(s).
\end{enumerate*}

\subsection{Lifecycle of Self-Descriptions}
\label{subsec:arch:lifecycle}
\begin{figure}
    \centering
\includegraphics[width=0.8\linewidth]{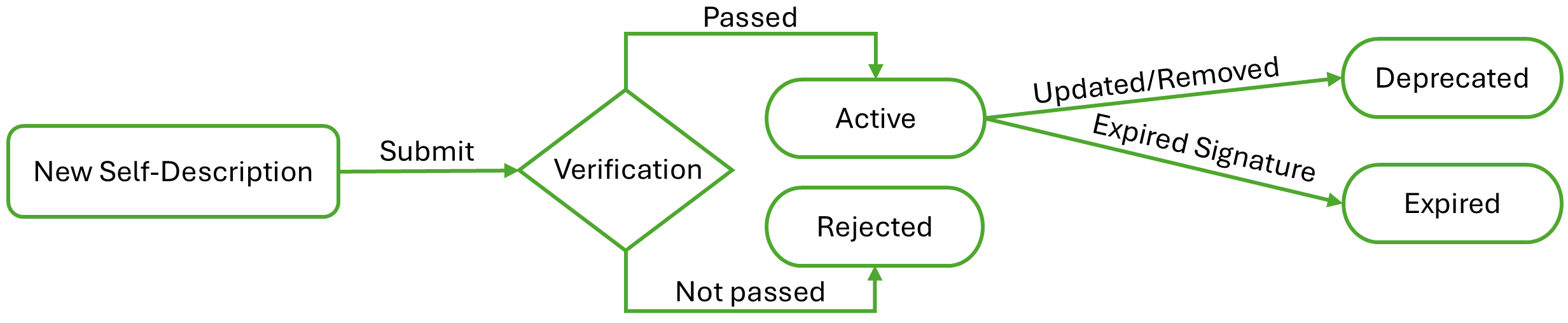}
    \caption{Steps in the lifecycle of a Self-Description}
    \label{fig:sd-lifecycle}
\end{figure}

Service offerings and their resources are subject to constant change. Therefore, their SDs need to be updated to reflect such changes. Due to the signatures, once created, SDs cannot be updated or changed. Instead, a new SD needs to be created and published, replacing the old one. Figure~\ref{fig:sd-lifecycle} shows the lifecycle.

Once submitted to the Catalogue, an SD is verified, as described in Section~\ref{subsec:arch:validation}. If the verification fails, the SD is not accepted and thereby rejected. After successful verification, the SD is in the \emph{active} state, which results in the claims of the SD being added to the SD graph and thus made accessible through the query endpoint (cf.\ Section~\ref{subsec:arch:query}). An SD and the corresponding claims are removed from the graph either if one of the signatures has \emph{expired}, it is explicitly revoked through the API or replaced by a new version (\emph{deprecated} state).
To determine if an SD is replacing an existing one (and to identify it), an identifier is needed. For this, the \textit{credentialSubject}'s \textit{@id}~\cite{W3C:VCDM} is used. Therefore, for each \textit{credentialSubject}, there is at most one SD in the active state (and thus a set of corresponding claims exposed through the query interface).

Of the Gaia-X Conceptual Model entities, \emph{Participant}, \emph{Service Offering}, \emph{Resources} and their subclasses are relevant for the Catalogue.  Other data structures such as a Participant's postal address do not occur stand-alone but only within an SD. This life-cycle management component satisfies Requirement~\ref{req:a}. While the XFSC Catalogue will only accept these three Gaia-X classes by default, this check can be disabled to also store other credentials. 

\subsection{Enforcing Consistency of Self-Descriptions through Schemas}%
\label{subsec:arch:schema-lifecycle}
Dataspace initiatives encourage their respective ontologies and shapes, as introduced in Section~\ref{sec:background}, to be extended inside federations to meet the domain specific requirements. Therefore, the Catalogue supports the submission of additional schemas.
Schemas define a consistent structure and semantics of SDs. We use ``schema'' as a generic term for either an ontology (OWL, defining classes and attributes) or a shapes graph (SHACL, defining validation constraints applied in the SD verification explained in Section~\ref{subsec:arch:validation}) or a controlled vocabulary (SKOS, defining reusable attribute values).
To manage schemas, the Catalogue assigns them a unique identifier. It expects ontologies to have an ontology IRI and controlled vocabularies to have a concept scheme URI. As SHACL does not specify a standard way of identifying a shapes graph, the Catalogue generates their IDs as SHA-256 hashes of their contents.
The Catalogue does not maintain a relationship between ontologies, shapes graphs and controlled vocabularies, as all of these may evolve independently.
The Catalogue always maintains a shapes graph formed as the union of all individual shapes graphs -- this is what SDs are verified against, satisfying Requirement~\ref{req:b}.

\subsection{Verification of the Self-Description}
\label{subsec:arch:validation}
\begin{figure}[!htb]
    \centering
    \includegraphics[width=0.8\linewidth]{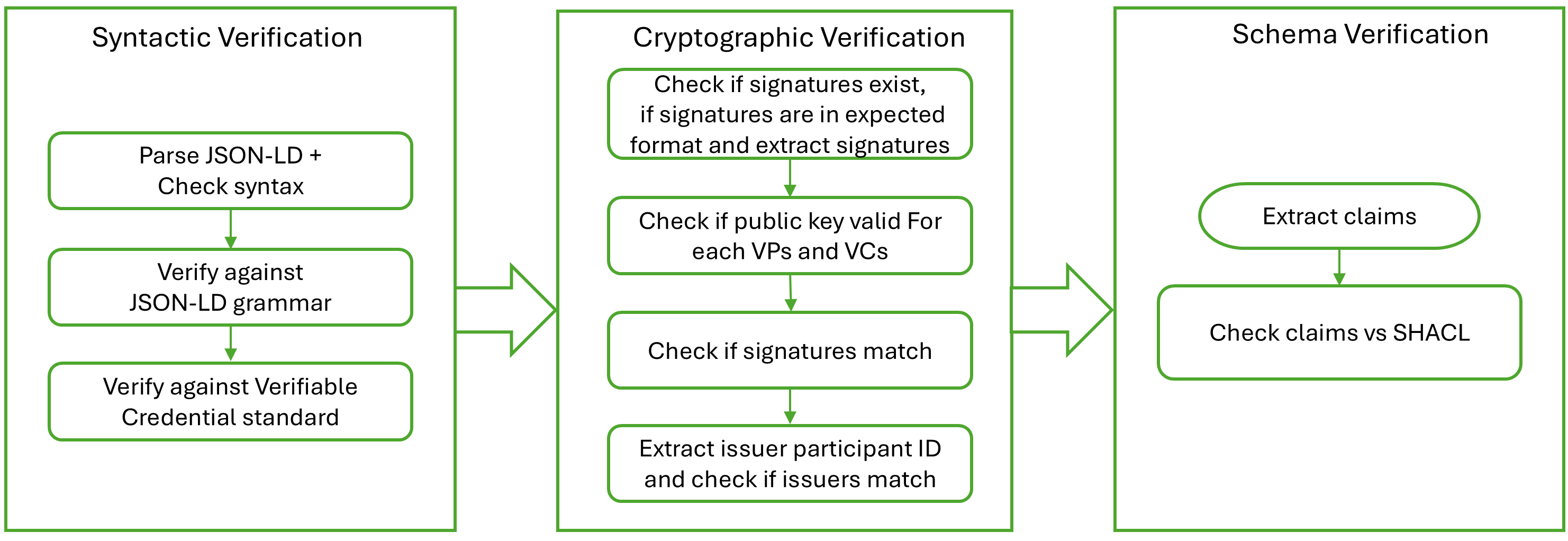}
    \caption{Flow of verification steps}
    \label{fig:verification_flow}
\end{figure}

To enforce SD validity (Requirements~\ref{req:a} and \ref{req:c}), SDs have to pass three verification steps before being accepted (see Figure~\ref{fig:verification_flow}).
In the first step, a simple \emph{syntax verification} checks whether the SD is valid JSON-LD and furthermore correctly implements the Verifiable Credentials Data Model~\cite{W3C:VCDM}.

In the second step, the \emph{cryptographic verification} is conducted, checking whether the VCs' and VPs' contents match the provided signatures, after URDNA 2015 normalization~\cite{W3C:RDFCanon} to ensure a system-agnostic interpretation.
The cryptographic verification initially iterates the VCs contained in the VP and individually checks their integrity. 
If these checks succeed, the VP is checked in the same way.
Additionally, the catalogue checks the participant ID to be consistent.

The third and last step is \emph{schema verification}. 
Here, a data graph built from all claims about SD subject is verified against the “union” shapes graph, i.e., against all constraints for semantic modeling that hold in the federation, according to the SHACL specification~\cite{ShapesConstraintLanguage}, and the schema specified by the Gaia-X Trust Framework which are pre-loaded in the Catalogue.
With a few exceptions, as mentioned in Section~\ref{sec:limitations}, the validation rules of the Gaia-X Trust Framework can be implemented in SHACL.
Currently, the following constraint types supported by SHACL 1.0 are used: %
\begin{enumerate*}\item \textit{sh:minCount} / \textit{sh:maxCount} defining the expected cardinality of an attribute,\item \textit{sh:minInclusive} / \textit{sh:maxInclusive} constraining an integer value to a certain range,\item \textit{sh:pattern} constraining a value to a certain pattern, \item \textit{sh:node} / \textit{sh:class} / \textit{sh:datatype} / \textit{sh:nodeKind} to constrain attribute values by type, and \item \textit{sh:in} to check whether a value is in a predefined list of possible values.\end{enumerate*}
In addition to SHACL validation, a credential subject is required to have a type supported by the Catalogue or an explicitly modeled subclass, i.e., \textit{Participant}, \textit{Service Offering}, or \textit{Resource}. While this check is enabled by default, it is possible to disable it, in order so support credential subjects of other class types.

\subsection{Querying the Self-Description Graph}
\label{subsec:arch:query}
To satisfy Requirement~\ref{req:d}, the Catalogue offers a powerful interface to store and query active SDs. 
During the specification phase in 2020/21, scalability concerns were raised about using SPARQL and RDF triple stores, i.e., handling queries on at least one million defined nodes, ten million properties, and ten million annotations~\cite{GXFS:CatalogueSpec}. 
Moreover, the maturity of RDF-star and SPARQL-star, as a then desired representation enabling the querying not only of SD claims but also the enclosing credentials was questioned, and query optimization challenges were identified~\cite{Hertweck.PerformanceSemanticQueries}.
To avoid scalability issues, the GXFS project chose openCypher and the labeled property graph (LPG) data model to implement this service. 
This implementation decision guarantees scalability and robust support in open-source databases. 
However, an interface with the semantic technologies used by other components is implemented.

\begin{figure}
    \centering
    \includegraphics[width=\linewidth]{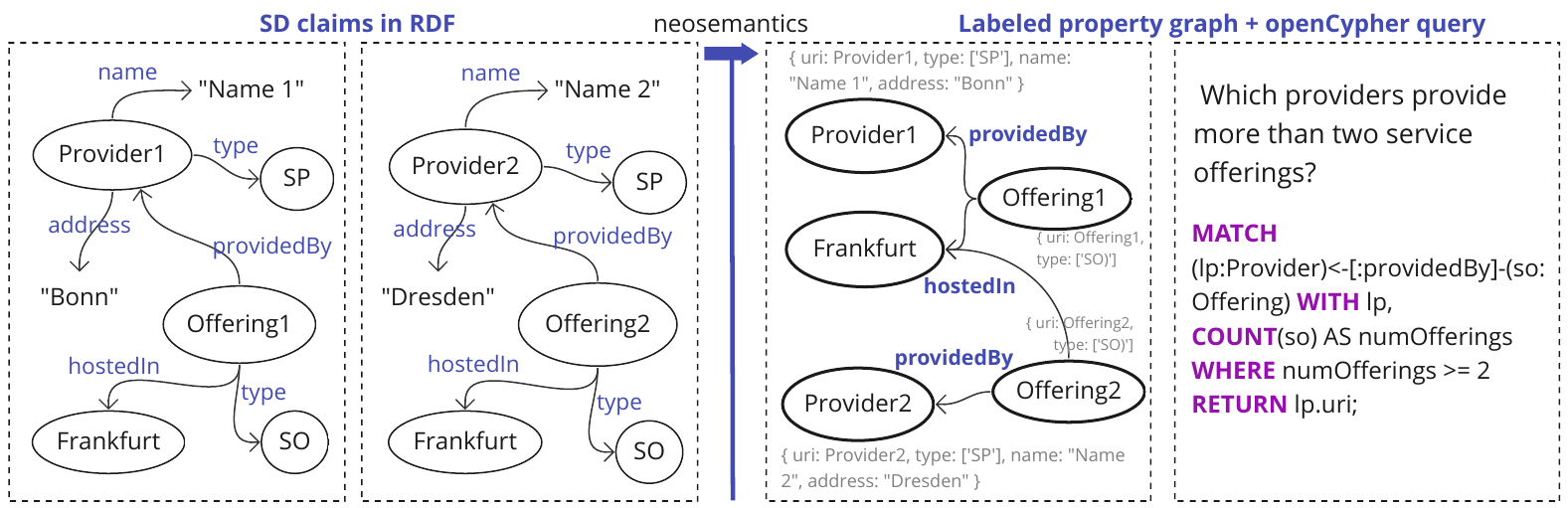}
    \caption{Example of transforming SD claims from RDF into a labeled property graph representation to run queries using openCypher.}
    \label{fig:transformation-rdf}
\end{figure}

We implement the catalogue's graph database in Neo4j using its ACID-compliant system. 
Neo4j's LPG data model does not directly match RDF, e.g., nodes and vertices are identified using database-internal identifiers that are not globally unique by URIs like RDF. 
Moreover, nodes and vertices can have properties and labels. Making RDF triples available in Neo4j requires transformation. 
We use the neosemantics plugin~\cite{neosemantics} to transform and store RDF data in Neo4j. 
Neosemantics provides a range of utility functions, e.g., importing and exporting data in different RDF serializations, including N-Triples and Turtle. 
Neosemantics is highly configurable through its Neo4j configuration node object, defined at graph initialization time. 
One critical setting in the context of the catalogue query service is the multi-valued (array-type) attribute to avoid overriding RDF properties that are not functional (in terms of OWL semantics) or type information in case multiple RDF types are given for a resource stored in Neo4j. 
For example, if a credential subject class is Provider and Vendor, both types should be kept. 
Once loaded into Neo4j, SDs are queried with openCypher. 
Figure~\ref{fig:transformation-rdf} shows querying providers with more than two service offerings.

\subsection{Integration with the EDC Connector}
 Technically, when one EDC is queried for its catalog, these offers are transmitted wrapped into a \texttt{dcat:Catalog}, serialized as JSON-LD. Its two relevant properties in this context are a \texttt{dcat:DataService} object, containing information on the queried connector, and an array of \texttt{dcat:Dataset} objects (or a single object), each of which represents an asset. Besides, the object contains a list of policies or a single policy with which the asset is offered, as described in section~\ref{sec:background}. The mapping of this structure to Gaia-X SDs is described below and is a core aspect of our integration.

Since the EDC is not compatible with our catalogue and is designed to communicate with other EDCs using the standardized \emph{Dataspace Protocol (DSP)}~\cite{dsp}, our overall architecture puts a so-called \textit{Broker EDC} in front of the XFSC Catalogue. For this, we make use of the extensible architecture of the EDC and implement an extension that acts as a facade towards the Catalogue. This way, the XFSC Catalogue can be used for powerful and efficient semantic queries while being transparent to the other EDCs implementation-wise. The Broker EDC periodically collects the other EDCs' catalogues, processes them and inserts them into the XFSC Catalogue. If another EDC wants to fetch the dataspace's catalogue, it can fetch the catalogue of the Broker EDC equally as it would query this EDC's own catalogue.
\begin{figure}[!htb]
    \centering
  \includesvg[width=0.7\linewidth,inkscapelatex=false]{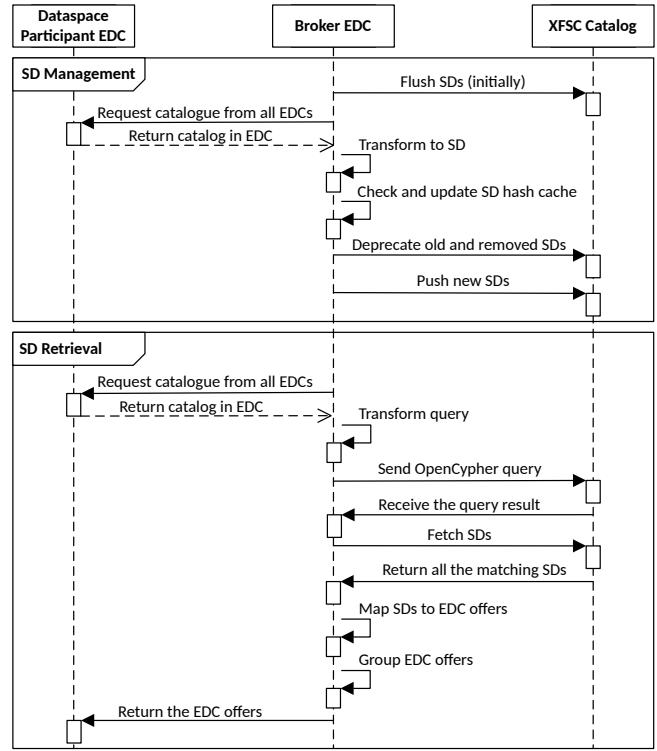}
    \caption{Binding XFSC Catalogue with EDC: Sequence diagram for integration workflow}
    \label{fig:edc_cat_maintenance}
\end{figure}

To maintain the XFSC Catalogue's content (see upper half of Fig.~\ref{fig:edc_cat_maintenance}), the catalogues of the dataspace's EDCs are periodically retrieved in the described DCAT structure. In a subsequent step, each catalogue is de-constructed into Gaia-X Resource SDs. Each asset is mapped to an individual SD, containing the metadata, the usage policies and also the connector metadata. Some inherent randomness in the SDs, such as the ordering of usage policies, is removed to obtain stable hashsums. The Broker EDC maintains a cache of hashes of SDs stored in the XFSC Catalogue. The list of the newly generated hashes is compared with the existing cache. SDs with hashes not yet in the cache are added to the catalogue. Hashes in the cache that do not correspond to a catalogue entry indicate that the respective SD is no longer present or has been updated. In both cases, the version in the XFSC Catalogue needs to be moved to the \emph{deprecated} state, as described in Section~\ref{subsec:arch:lifecycle}. The API calls for the XFSC Catalogue are generated by a modular software component following the facade pattern based on these two possible outcomes that induce changes to the catalogue.

The retrieval from the XFSC catalogue works as follows (see lower half of Fig.~\ref{fig:edc_cat_maintenance}): We patch the Broker EDC's standard process of catalogue retrieval such that we handle the construction of the returned catalogue ourselves. For filtering catalogues on retrieval, the EDC supports a generic list of (left operand, operator, right operand)-triples. Our EDC extension contains a mapper that translates these triples into openCypher queries, considering the overall SD structure and the structure of the graph constructed as described in Section~\ref{subsec:arch:query}. Based on the \emph{credentialSubject} IDs that are obtained from the knowledge graph, the respective SDs are retrieved and re-mapped to EDC catalogues. To obtain a list of catalogues, one per dataspace participant, we group the SDs by the participant stated in the \texttt{dcat:DataService}. During translation, the associated assets with their usage policies are then merged into a single list of \texttt{dcat:Dataset} objects and returned to the querying EDC.

\section{Implementation and Demo}
\label{sec:implementation}
The implementation of the XFSC Catalogue strictly followed a comprehensive software requirements specification~\cite{GXFS:CatalogueSpec}. It is implemented as a web service providing its functions through a REST API. It is designed as a scalable, container-based application that can be easily deployed into any cloud infrastructure. The complete architecture and design decisions are documented publicly~\cite{CAT:ArchitectureDocument}.

The core is implemented in Java using the Spring Boot framework. The graph database for storing the SDs' claims and making them queryable is realized by Neo4j and neosemantics (cf.\ Section~\ref{subsec:arch:query}). Submitted SD and schema files as well as administrative information, e.g., lifecycle state of SDs or validity periods of signatures are stored in a relational database (PostgreSQL). The authentication and authorization workflow is based on the OpenID Connect standard~\cite{OIDC} and uses JSON Web Tokens~\cite{RFC7519}. It is a dedicated component, which allows exchanging the implementation (e.g., switching from Keycloak in our implementation to the XFSC Authentication \& Authorization component).

Besides the REST API, a simple web-based interface exposes user management, verification and query functions.
Realistic exercises to try with the web interface and, via the Postman client, with the REST API, were prepared for a community workshop (cf.\ Section~\ref{sec:adoption}) and have been recorded as videos: \begin{enumerate*}\item submitting an SD that requires a non-standard schema to validate,\item managing the lifecycle of participants and schemas, and \item running queries~\cite{fc-exercises}.  \end{enumerate*}

We developed the EDC integration based on EDC version 0.2.1, albeit version 0.10.1 is available at the time of writing, due to the broad adoption of version 0.2.1 and breaking changes between them. To account for this and to future-proof our implementation, we have built the concerned features in a modular fashion for easy replacement of especially the SD $\leftrightarrow$ EDC catalogue mapping.

\section{Limitations}
\label{sec:limitations}
Since the realization of the Catalogue followed a strictly linear development cycle (completed specification phase before starting the implementation), some limitations were recognized and necessary during the implementation.

\noindent\textbf{Standards conformance:} In general, VPs have a flexible structure to describe entities~\cite{W3C:VCDM}.  Gaia-X foresees the case that a single SD describes multiple entities, i.e., that its VCs address multiple credential subjects~\cite{GaiaX:ICAM}. In contrast, our Catalogue only accepts a SD if all its VCs address the same subject. This is required to ensure manageable update and replacement mechanisms for SDs and for the claims in the graph database. The credential subject's URI is used as an identifier of the SD. A URI (\textit{@id}) is generally required in Gaia-X~\cite{GaiaX:ICAM}, while it is optional according to W3C~\cite{W3C:VCDM}. In summary, the Catalogue only accepts a subset of all valid SDs, which in turn is a subset of all valid VPs.

\noindent\textbf{Credential assembly:} The Catalogue currently only accepts self-contained SDs. In practice, it may be convenient to keep various VCs, e.g., in a wallet, and only present them on request (assembling a VP on the fly) or to selectively disclose individual credentials or even individual claims, i.e., the \emph{declaration} they initially publish to the Catalogue would not include such sensitive details but only references to them, which would be disclosed to certain requesters only. Support for an advanced workflow was worked on as part of the XFSC Integration project~\cite{GXFS-Integration}. The concept is that providers may decentrally keep credentials in local SD storages, which the Catalogue crawls via the Decentralized Identifiers (DIDs) that identify the credential subjects, to finally obtain a full VP for the given purpose. While no updated technical portal specification containing this concept was published under the XFSC project, these ideas will partially be addressed in the successor project FACIS (Federation Architecture for Composition of Infrastructure Services). %

\noindent\textbf{Variety of validation rules:} Since the Catalogue validates SDs against the union of all shapes graphs, only one node shape applies to any given subject.  This makes sense for the classes defined by the Gaia-X working groups, since the constraints on their attributes have been agreed upon by consensus, and no two contradicting shapes should exist for the same Gaia-X class. However, this limits the reuse of standard ontologies for general concepts, e.g., using the vCard ontology~\cite{vCardRDF} for postal addresses. When there is only one node shape targeting the \textit{vcard:Address} class, it is not straightforward to define different constraints per context (e.g., requiring all address fields to be present for certain types of data center, whereas the current SHACL shapes stored in the Gaia-X Registry only require defining a country code for a Participant~\cite{gaiax-registry}). This limitation can be worked around by adding more explicit context information to the shapes (e.g., not by broadly using \textit{sh:targetClass}, but more specifically using \textit{sh:targetObjectsOf} or by defining shapes specific to the object of a property via \textit{sh:property/sh:node}), but doing so may require rewriting existing shapes.
Further, the Trust Framework currently defines one validation rule that cannot be expressed in SHACL (``if[, for a \textit{LegalPerson},] several \textit{[registration]Number}s are provided, the information provided by each number must be consistent''~\cite{GaiaX:TrustFramework}) and thus not validated by the Catalogue; in federations, many such rules might be expected.

\noindent\textbf{Integration with the EDC:} Building upon EDC version 0.2.1 is a significant limitation, though without alternatives due to the German Culture Dataspace project's external software dependencies. The modular architecture with clearly defined interfaces for the individual components aims at a low-effort transition to newer EDC versions. Besides, the crawling-based architecture of our approach does not necessitate any actions from the dataspace participants, allowing the XFSC Catalogue to be used as a drop-in replacement for existing catalogue implementations, as compared to a provider-push architecture. As a drawback, the offering participant's connector never learns about validation and verification failures and might wonder why an offer does not appear in the catalogue.

\noindent\textbf{Federation service integration:} The components in the XFSC Toolbox evolved individually; their integration is still work in progress~\cite{GXFS-Integration}. Integration of the Catalogue with the XFSC Authentication \& Authorization (cf.\ Section~\ref{sec:implementation}) and credential management components (see above) is of particular relevance.

\section{Evaluation}
\label{sec:evaluation}
The XFSC Catalogue was evaluated against the functional and quality requirements stated in its specification, totalling 151. Of these, 122 were successfully met, with the remaining 29 all being optional requirements~\cite{GXFS:CatalogueSpec}. %
To ensure high software quality, all implemented components are covered by unit tests covering their basic functionality, which run automatically via continuous integration (CI).
In addition, extensive functional, performance and penetration tests were done~\cite{fc-testresults}.

\textbf{Functional tests}
were based on a deployed Catalogue instance, in a production-like environment, testing the REST API endpoint of each component with Postman.
The Postman collection used for testing contains all requests to the FC API and can be found in the Pre acceptance testing repository, which also contains videos of end-to-end testing steps for each functional block, proving how the original acceptance criteria are implemented~\cite{fc-testing}. Separate tests were done regarding the non-functional performance and scalability criteria, as well as the requirements for application security.

\textbf{Performance and scalability} were major quality requirements, as the specification anticipated a high number of SDs and API requests.
To prove this, load tests were executed using Gatling~\cite{Gatling}, testing the two major scenarios: \begin{enumerate*}\item \emph{Self-Description Management}: adding, retrieval, revocation and deletion of SDs. \item \emph{Query processing}: executing several openCypher queries through the query endpoint.\end{enumerate*}
Scalability was also tested by deploying multiple instances of the Catalogue.
To prepare a big number of valid SDs, two tools were developed~\cite{fc-tools}: \begin{enumerate*}\item \textbf{Generator}: to generate SD files, \item \textbf{Signer}: to sign the generated SDs.\end{enumerate*}
While the Generator was designed to create a large number of flat claims for performance testing, more complex and more realistic SDs were created manually to be used for technical demonstrations and exercises (cf.\ Section~\ref{sec:implementation}).
\textbf{Penetration tests} were conducted by an independent security testing team. Identified vulnerabilities were mainly related to information disclosure and were mitigated by hardening the Catalogue's Ingress configuration.

As development will continue as an open source project, \textbf{improvements and extensions of the testing process} are planned: automating the functional API tests and including them into the automated CI/CD pipeline, or extending the performance tests to millions of active SDs (and thus tens of millions of nodes in the graph database).
In addition to the formal evaluation of the Catalogue implementation, the test of its practical applicability through adoption has begun.

\section{Adoption and Sustainability}
\label{sec:adoption}
The adoption of XFSC and/or alternative Gaia-X federation service approaches (cf.\ Section~\ref{sec:related}) can be seen in some of the 16 Gaia-X Lighthouse projects~\cite{GaiaX:Lighthouses}, and 11  projects of the German economy ministry's funding competition ``Innovative and practical applications and data spaces in the Gaia-X digital ecosystem''~\cite{BMWK:Foerderwettbewerb}, based on the requirements of their respective use cases. For example, some projects of the Gaia-X 4 Future Mobility project family use the XFSC Catalogue in their dataspace implementations. One of the projects, Gaia-X 4 PLC-AAD (Product Life Cycle -- Across Automated Driving)  integrated the XFSC Catalogue with their claim-compliance-provider~\cite{claim-compliance-provider}, enabling users to easily create claims about their services, check their compliance with Gaia-X and store them in the XFSC Catalogue. Other projects, such as Gaia-X 4 ROMS (Support and Remote-Operation of automated and interconnected Mobility Services) also state that the XFSC Catalogue is used on their homepage~\cite{gx4roms}. In the Gaia-X 4 AMS (Advanced Mobility Services) project the XFSC Catalogue was recently deployed and will be used to store service offering credentials~\cite{ams-catalogue}.

Several Gaia-X implementation projects require sovereign data exchange and thus use the EDC, but require more comprehensive federation service functionality than the EDC provides itself.  This yields the motivation for an integration of the two approaches.
Beyond dataspaces built by individual projects, larger-scale initiatives such as Simpl, the Smart Middleware Platform for Cloud-to-Edge Federations and Data Spaces procured by the European Commission, which is broader in scope than Gaia-X, offer potential for adoption. 

The integration of the XFSC Catalogue with the EDC has been done with immediate adoption within the German Culture Dataspace in mind. This dataspace is currently transitioning from a demonstrator phase to regular operation, establishing its own governance body and technical operator. %
It comprises more than ten participating institutions with individual connectors. The catalogue has been deployed in this context as a drop-in replacement and is performing well. The enhanced semantic capabilities in comparison to the EDC's catalogue features allow for specific queries. The combination of heterogeneous, rich metadata in the cultural sector and the comparatively lower degree of digital maturity of its actors compared to other sectors makes the XFSC Catalogue particularly well suited in that environment, being used seamlessly via the EDC and making the rich metadata usable.

The dissemination of the XFSC Catalogue was done through the GXFS (later XFSC) Tech Workshops~\cite{gxfs}. Six out of eight workshops had a dedicated Catalogue track or session; a further focus topic was Identity \& Trust.  The workshops mainly targeted projects funded by the German economy ministry. The last workshop, held in 11/2024 in Madrid, was the first one actively targeting a non-German audience: the Spanish Gaia-X hub. In addition to the documentation of these workshops, new users can get support from the active community around the GitLab repository~\cite{fc-service} to document and work on upcoming issues. As of December 2024, 254 issues have been opened, of which 226 were closed (41 of these in the last 9 months). Overall, 309 merge requests were successfully merged (30 of these in the last 9 months).

\section{Conclusion and Future Work}
\label{sec:conclusion}
We presented the XFSC Catalogue as a directory for Gaia-X services and their providers, integrated with a widely used dataspace connector and in operation in a German Culture dataspace, a lighthouse project within Germany's national Digital Strategy. Developed within the scope of Gaia-X, its application is not limited to Gaia-X. The catalogue contains Self-Descriptions (SDs) to ensure interpretability, comparability and transparency, based on broadly adopted Semantic Web standards. Schemas are used to ensure a consistent and well-defined representation of the metadata. Cryptographic signatures are used to ensure integrity and authenticity in this federated and distributed environment and thus foster trust in providers' claims.

During implementation, it became clear that, when adopting the W3C standards for Gaia-X, there is room for interpretation. Although the Gaia-X and GXFS specifications have narrowed this down, the implementation points out further demands. Some of the limitations pointed out in Section~\ref{sec:limitations} can be removed by clarifying the imprecise points in subsequent specifications.
At the same time, the Gaia-X specifications are still evolving towards a cleaner grounding in established standards, as it has happened with the structuring of Gaia-X Credentials its alignment with the more general ISO Conformity Assessment standard~\cite{ISO17000}.%

In addition, the query interface of the catalogue technically provides a GraphQL endpoint. In interaction with the XFSC portal, a more user-friendly interface can be established.

For sustained adoption of the catalogue, enhancing its integration with the EDC Connector to support newer versions of the latter will be crucial. At the time of writing, the new ``Loire'' release of Gaia-X specifications had just become available~\cite{gaiax-docs}, including breaking changes with respect to the SDs -- thus, also compatibility with this new release will be crucial.

\paragraph{Acknowledgments:} The implementation was carried out in a contract for the eco Association of the Internet Industry, resulting from a tender they had run in the scope of the Gaia-X Federation Services project funded by the German Federal Ministry for Economic Affairs and Climate Action (BMWK, grant 13I40V007A). The integration with the EDC was part of the project ``Datenraum Kultur'', funded by the German Federal Government Commissioner for Culture and the Media (BKM) under grant number 2522DIG012. We express our gratitude to Nikhil Acharya, Julius Pfrommer, and Patrick Westphal who have significantly contributed to the implementation and a prior version of this manuscript.

\paragraph{Resource availability statement:} \noindent\textbf{Resource Type:} Software Service,\\\textbf{License:} Apache 2.0, \textbf{Permanent URL:} \url{https://w3id.org/xfsc/catalogue/} (cite as~\cite{GXFS-Catalogue}.%
)
\newpage
\printbibliography
\end{document}